\documentclass[final]{aipproc}

\layoutstyle{6x9}
\def\lsim{\mathrel{\rlap{\lower4pt\hbox{\hskip1pt$\sim$}}
    \raise1pt\hbox{$<$}}}         
\def\gsim{\mathrel{\rlap{\lower4pt\hbox{\hskip1pt$\sim$}}
    \raise1pt\hbox{$>$}}}         

\begin{document}

\title{Nuclear Astrophysics}


\author{W. C. Haxton}{
  address={Institute for Nuclear Theory and Department of Physics, \\
  University of Washington, Seattle, WA 98195}
}

\begin{abstract}
 I review progress that has been made in nuclear astrophysics over the
 past few years and summarize some of the questions that remain.  Topics
 selected include solar neutrinos, supernovae (the explosion and associated nucleosynthesis),
 laboratory astrophysics, and neutron star structure.
\end{abstract}

\maketitle


\section{Solar Neutrinos and the CNO Cycle}

 Perhaps no problem in nuclear astrophysics has drawn
 more attention than solar neutrinos and the associated discoveries of neutrino mass and
 mixing, the first experimental evidence for physics beyond the minimal standard model.
 These results grew out of careful solar modeling, patient laboratory measurements of pp-chain
 nuclear cross sections, and heroic efforts on new experiments.  The detectors that have
 come on line in the past decade and resolved the solar neutrino puzzle, SNO \cite{SNO} and
 Super-Kamiokande \cite{SuperK}, are remarkable instruments.
 
 But much remains to be done.  Direct event-by-event measurements are so far
 limited to the high-energy $^8$B neutrinos, a 0.01\% branch in the sun.  There are important
 reasons for extending such measurements to lower energies: \\
 $\bullet$ No phenomenon directly
 connected with matter-enhanced oscillations, such as spectral distortions or day-night effects
 due to the earth's matter, has been observed.   While indirect evidence of 
 spectral distortions comes from the comparison of  GALLEX/GNO/SAGE \cite{Gallex} integrated
 rates with SNO 
 and Super-Kamiokande results, a direct measurement of changes in  $P_{\nu_e}(E_\nu)$ as a function
 of $E_\nu$ is lacking.
 The parameters deduced from SNO, Super-Kamiokande, and earlier radiochemical experiments
 predict that low-energy pp and $^7$Be neutrinos experience only vacuum oscillations,
 as the matter contribution to the effective neutrino mass
 \begin{equation} 
 m_{\nu_e}^{2~eff} = 4 E_\nu \sqrt{2} G_F \rho_e(x)
 \end{equation}
 is insufficient to cause a level crossing for low-energy $\nu$s.  The survival probability for these
 sources is almost twice that for the high-energy portion of the $^8$B $\nu_e$ flux.  
 There is some chance that current experiments will be able to see variations in $P_{\nu_e}$.
 SNO-III (use of $^3$He neutron counters) may be able to operate at a lower threshold
 due to improvements in the charge-current(CC)/neutral current(NC) event separation.
 KamLAND \cite{KamLAND} will be attempting a $^7$Be measurement (though this is a difficult task at 
 Kamioka depths), and Borexino \cite{Borexino} may soon turn on at Gran Sasso.  \\
 $\bullet$ Measurement of the principal component of the solar flux, pp neutrinos,
 would isolate one of nature's best calibrated sources of $\nu_e$s.
 In part because SNO and Super-Kamiokande measurements of $^8$B neutrinos have
 constrained the temperature of the solar core, the standard solar model (SSM)
 uncertainty in the
 pp flux is $\lsim$ 1\%.  For this reason pp experiments could improve the accuracy
 of mixing angle ($\theta_{12}$) determinations.  New detectors,
 such as CLEAN \cite{CLEAN} (NC scattering in liquid neon) and 
 LENS \cite{LENS} (CC scattering in In), are being
 developed with this goal in mind.\\
 $\bullet$  But perhaps the most intriguing solar neutrinos are those from H-burning through
 the CNO cycle, which contributes to solar energy generation at about the 1\% level. 
 The CNO cycle is the dominant
 mechanism for stellar energy generation in heavier main-sequence stars, including 
 the very massive popIII stars that formed and produced "first light" (and which turned on the CNO
 cycle by first synthesizing carbon in the triple-alpha process) \cite{firststars}.  Indeed, nuclear
 astrophysics arguably began with Bethe's search for the mechanism that could account for
 energy generation in massive main-sequence stars.
 
 Because the CNO $\nu$ flux is proportional to core metallicity Z, a measurement would test
 this important parameter.
 It is believed that the sun was highly convective as it entered the main sequence, and
 thus thoroughly mixed.  A plausible consequence --
 that the zero-age solar core metallicity Z can be
 equated to today's surface abundance -- is a key postulate of the SSM \cite{bahcallSSM}.
 SSM predictions are sensitive to Z because of the importance of
 free-bound atomic transitions to the solar opacity.  Furthermore, metals 
 can have interesting evolutionary consequences by driving 
 out-of-equilibrium burning of CNO elements in the early sun.  Such burning is thought to 
 have powered an early convective solar core for $\sim$ 10$^8$ years.
 
 The sun's metallicity also controls the location of the radiative/convective
 boundary: a shift to lower Z makes radiative transport in the interior more efficient, and thus
 leads to a shallower convective zone.  Recently
 improved 3D models of the solar atmosphere have altered the interpretation of
 photospheric absorption lines, lowering deduced metal abundances by 30\% \cite{3Datmos}.  This has
 destroyed the previous agreement between SSM predictions and helioseismology: the
 acoustic modes are quite sensitive to the depth of the convective zone.  This is yet another
 reason for attempting an independent measurement of the sun's interior Z.
 
 \section{Understanding Core-Collapse Supernovae \cite{corecollapse}}
 Massive stars evolve quickly because their rates of energy generation must be sufficient to
 sustain hydrostatic equilibrium in deep gravitational potentials.  They progress through a
 series of burning stages and contractions until the final explosive stage of Si burning,
 where the iron-group ashes that are produced are inert (as these are nature's most tightly
 bound nuclei).  When the iron core grows to the Chandrasekhar mass, it begins a collapse
 that proceeds on
 nearly free-fall timescales, as the ability of the electron gas to sustain the star is further eroded by
 the emission of electron-capture
 neutrinos and by the excitation of nuclear states, processes taking energy from the
 gas.  Often core collapse leads to ejection of the star's mantle - a supernova (SN) --
 and a spectacular optical display.
 
 The fundamental problem in understanding the explosion is energy transfer:
 Because all volume elements in the pre-collapse star are bound gravitationally, the
 collapse mechanism must preferentially transfer released energy to the star's mantle,
 to account for its ejection.  One potential mechanism for energy transfer is the shock wave
 that forms at the edge of the proto-neutron star, the ``sonic point'' where the infall velocity
 matches the local sound velocity.  Here pressure waves, initially produced when infalling shells of
 matter rebound at super-nuclear densities due to nuclear incompressibility,
 collect after traveling outward from the collapsing
 star's dense center.  When the sonic point itself comes to rest, the shock wave is
 released to travel through the star's outer iron core and mantle.  As it does so, it melts the iron into a
 nucleon and alpha gas.  This melting (which requires $\sim$ 8 MeV/nucleon)
 and neutrino emission from the matter robs the shock of its energy, typically stalling the
 shock 200 to 300 km from the star's center.  It becomes an accretion shock: the infalling
 matter crossing the shock front both delivers energy and extracts it, due to nuclear
 melting, in about equal measures.  The shock wave pauses.
 
 Thus other energy transfer mechanisms must operate.  The hot, protoneutron star
 radiates neutrinos, which can scatter off the nucleons in the nucleon gas left in the 
 shock wave's wake.  The relatively strong CC reactions heat the gas, building up 
 pressure that, after a second or so, helps to push the shock wave outward again.  However
 this ``delayed mechanism'' also fails to succeed in 1D calculations -- the regions of the star
 over which the neutrinos can do useful work and the time over which that work is performed
 are limited.  Matter near the protoneutron star surface, where the neutrino flux is intense,
 can be heated efficiently.   But once T $\sim$ 2 MeV is reached,  the matter re-radiates
 neutrinos, effectively halting net heating.   Matter further out may not be appreciably heated, given
 the lower neutrino flux and the short period, $\sim$ 1s, relevant to the explosion mechanism.
 
 This has led to the recognition that the explosion mechanism may be fundamentally multi-D:
 1-D models may place artificial limits on the effectiveness of neutrino heating.  In 2- or 3-D,
 convection can sweep hot, neutrino-heated matter to large radii, cooling it before it can
 re-radiate neutrinos.  This matter is replaced by descending cool material, allowing this matter to be
 efficiently heated.  It is clear that convection allows more energy to be deposited in and
 retained by the star's mantle.
 
 The supernova mechanism is a grand-challenge numerical problem, an important but complex process
 that requires realistic modeling of  magneto-hydrodynamics, shock wave propagation, and
 neutrino diffusion (including transport of energy and lepton number) in multi-D.  Important
 progress has been made.  Improvements in the
 modeling of electron capture and $\beta$-decay in the early stages of core collapse have
 helped refine estimates of the core mass \cite{langanke}.  
 Simulations in 1-D with full Boltzmann transport and with three-flavor neutrino
 transport have been
 performed \cite{mezz}.  Several groups have explored 2- and 3-D models, though necessarily with 
 simplified transport assumptions \cite{3D}.  New instabilities have been identified, 
 such as
 a spherical shock instability that influences matter infall and leads to neutron star spin-up \cite{spin}.
 Some of this progress is due to the the Department of Energy's program for Scientific
 Discovery through Advanced Computing, which has made core-collapse SN modeling one
 of its terascale computing applications.
 
 But many challenges remain.  Modelers have not yet identified a robust explosion mechanism,
 perhaps because it is not yet possible to combine multi-D modeling
 with state-of-the-art microphysics, neutrino transport, neutrino flavor violation,
 magnetohydrodynamics, general
 relativity, and rotation.  There are significant uncertainties in the input
 physics, including the nuclear equation of state (EOS) at sub- and super-nuclear densities, 
 the effects of trapped neutrinos on the MSW potential governing oscillations, and the nuclear
 physics of finite-temperature reactions.  There are also important
 challenges for experiment, such as detecting all neutrino flavors during the next galactic
 SN and determining nuclear reaction cross sections important to SN nucleosynthesis --
 topics for the next two sections of this talk.
 
 \section{Nucleosynthesis and Supernova Neutrinos}
 A problem likely connected with core-collapse SN is the origin of r-process nuclei.
 The r- or rapid-neutron-capture-process is thought to produce about half of the nuclei heavier than iron, including all of the transuranics.  Candidate astrophysical sites must be explosive,
 such as core-collapse SN and neutron-star merges, to produce the requisite
 temperatures and neutron fluences to drive the r-process.
 
 We have learned a great deal about the r-process in the past few years. 
 Abundance data from old metal-poor stars have revealed an r-process distribution for
 (Z,A) $\gsim$ (56,130) that matches solar abundances \cite{cowan}.  
 As it is believed the surfaces of such
 stars would have been enriched by very few nearby SN, this suggests that the
 mechanism that produces high-mass r-process nuclei is a signature event,
 operating now as it did in early
 stars.  In contrast, the distribution of lighter r-process nuclei
 varies considerably.  The observations, which include chronometer results suggesting 
 different frequencies for events responsible for synthesizing low- and high-mass nuclei,
 may indicate two or more classes of sites for the r-process, or perhaps a single class
 where the production of low-mass nuclei is highly irregular \cite{qian}.
 
 The r-process is a mechanism where nuclei are synthesized under explosive, neutron-rich
 conditions.  Nuclear equilibrium is maintain by (n,$\gamma) \leftrightarrow (\gamma$,n)
 reactions, rather than the $\beta$-decay equilibrium that defines the valley of stability for
 ordinary nuclei.  The neutron-rich nuclei that exist under such conditions can capture a 
 neutron whenever $\beta$-decay occurs, n $\rightarrow$ p, as this opens up the needed 
 hole for $(n,\gamma)$.  Thus the rate of nucleosynthesis is controlled by $\beta$-decay.  This
 rate slows near closed shells because repeated $\beta$-decays are needed to overcome
 shell gaps.  Thus mass piles up at these points,
 leading to abundance peaks near A $\sim$ 130, 195.  After the explosion producing the
 neutrons ceases, the unstable parent nuclei decay ($n \rightarrow p + e^- + \bar{\nu}_e$) 
 to the valley of stability, thereby
 producing daughters with N/Z characteristic of stable nuclei.
 
 Of the two sites most often discussed, the expanding, cooling, $\nu$-driven winds of a Type II
 SN and the decompressing n-rich matter from neutron star mergers, data now favor
 the former.  The two mechanisms involve comparable explosion energies and thus would 
 mix newly synthesized elements into comparable volumes in the interstellar medium.
 Yet neutron star mergers are roughly 1\% as frequent as SNIIs.  Thus, to be
 the dominant site for the r-process, the mass produced per neutron star merger
 must be 100 times larger.
 This would be reflected in very large enrichments of isotopes like Eu in early, metal-poor
 stars -- but such enrichments are not seen.
 
 The hot nucleon gas that is left in the wake of the shock, then later blown off the star by the neutrino 
 wind, is thought to be the SN r-process site \cite{woosley}.  
 This n-rich material, on expanding and cooling, undergoes an $\alpha$
 freezeout and then $\alpha$ capture reactions to produce some heavy seed nuclei
 of intermediate mass.
 Excess neutrons are left over.  If conditions can be arranged so that the resulting n/heavy seed
 ratio is $\sim$ 100, the neutrons will be able to drive an r-process  up to the A $\sim$ 195
 mass peak.
 
 Our lack of understanding of the SNII explosion mechanism leaves open a wide parameter
 space for the r-process.  We do not know what conditions -- temperature T, density $\rho$,
 and entropy S -- describe the trajectories of volume elements ejected from the star.  We do
 not know the neutrino fluence through these volumes: this is crucial because neutrinos are
 needed to lift the ejected material off the star, yet neutrino reactions can destroy free
 neutrons and greatly decrease the neutron/seed ratio, thereby making an r-process
 impossible \cite{fullermeyer}.
 
 New neutrino discoveries have added yet another layer of complexity.
 The parameter governing flavor conversion
 in the relevant regions of the star,  $\theta_{13}$, has not yet been measured.  But unless it
 is very small, $ \lsim 10^{-4}$, an adiabatic level crossing
 will occur.   The naive position of this crossing is near the base of the carbon zone,
 at a density of about 10$^4$ g/cm$^3$.  But this estimate ignores the contribution of the
 neutrino background to the MSW potential.  It appears, from calculations done to date, that
 neutrino self-interactions will move the flavor conversion point much deeper into the
 star \cite{fullernu}.  This could have very important consequences, as the conversion might affect
 both the explosion and nucleosynthesis in important ways.  The reason has to do with the
 higher temperatures characterizing heavy-flavor neutrinos, reflecting their
 weaker matter couplings.  Consequently, a $\nu_e \leftrightarrow \nu_\tau$ oscillation effectively
 heats the $\nu_e$ spectrum, enhancing the CC interactions that
 govern the n/p ratio and dominate energy transfer from neutrinos to the matter.
 Flavor conversion could be altered by a variety of effects only poorly understood and
 modeled, including the density inhomogeneities that are produced by passage of the
 shock wave and the high-velocity matter flows induced by convection \cite{haxtonfuller}.  
 (The MSW potential
 also includes a coupling between the neutrino's velocity vector and weak currents.)
 
 Another topic attracting interest is the possibility of detecting the SN background $\nu$s,
 the ``relic'' neutrinos that have accumulated during the entire epoch of star formation \cite{lunardi}.
 Such a measurement should be possible in future megadetectors: Super-Kamiokande has already
 established a limit that is within about a factor of three of theoretical predictions \cite{SKrelic}.
 Theorists have been busy considering the star formation rate, the stellar mass ranges that
 might lead to core collapse, and the spectral distortions due to redshift.
 
 SN neutrinos not only govern the conditions for the r-process, but can also directly
 synthesize new nuclei through the neutrino process.  Even far out in the mantle, such as the
 Ne zone, the neutrino fluence is so high that an appreciable fraction of nuclei, $\gsim$ 0.1\%
 typically, undergo inelastic reactions.   Often this leads to excitation of giant resonances 
 followed by the emission of a proton or neutron, and thus the production of a daughter 
 isotope with mass A-1.  In the past few years some important progress in neutrino process
 calculations has occurred.  Heger et al. extended existing network calculations to heavier 
 nuclei, demonstrating in the process that the rare isotopes La and Ta are likely neutrino
 process products \cite{heger}.  La production is dominated by CC reactions off $^{138}$Ba, a departure
 from the usual pattern of NC dominance
 of neutrino nucleosynthesis.  Frohlich et al. \cite{frohlich}
 and Pruet et al. \cite{pruet} demonstrated that neutrino reactions on
 protons could produce a neutron flux sufficient to synthesize p-process nuclei, another example of 
 a CC process.  These CC $\nu$-process channels could be altered by oscillations.
 
 \section{Concordance in Micro-Macro Physics}
 One of the remarkable recent achievements in astrophysics is the agreement between 
 the values determined for $\eta$, the baryon/photon ratio, from Big-Bang nucleosynthesis
 (BBN) and from studies of the microwave background and large-scale structure formation \cite{eta}.
 The former involves nuclear microphysics within the first three minutes of the birth of the
 universe.  The latter reflects the large-scale behavior of matter -- specifically the baryonic
 matter-radiation interactions that retard baryon infall into dark-matter halos -- at the time
 of recombination, about 400,000 years after BBN.  Thus the agreement ties together two
 phenomena that occurred at very different times and involved very different scales, one
 nuclear and one cosmological.
 
 In the future we can look forward to two similar concordance tests.  A nonzero $\eta$ requires
 an excess of baryons over antibaryons in the early universe.  One of the
 three conditions necessary for baryogenesis
 is CP violation.  But while CP violation has been observed among the quarks, theorists have 
 concluded that the known strength is insufficient to explain $\eta$.
 
 However recent discoveries in neutrino physics suggest a resolution of this puzzle.
 The three-generation neutrino mass matrix has three CP-violating phases, one (Dirac) that
 could influence long-baseline neutrino oscillation experiments and two (Majorana) that
 could play a role in low-energy processes like neutrinoless $\beta \beta$-decay.   The quantity
 that could be tested in neutrino oscillations, e.g., by comparing the probabilities for
 $\nu_\mu \rightarrow \nu_\tau$ with $\bar{\nu}_\mu \rightarrow \bar{\nu}_\tau$ while accounting
 for the effects of matter on this difference, involves a product of neutrino mixing parameters,
 all of which are known to be large except for $\theta_{13}$ (not yet measured) and the
 CP phase.  But unless $\theta_{13}$ is very small (e.g., $\lsim 10^{-4}$), there are
 accelerator strategies that will be able to constrain the CP phase \cite{neutrinoCP}.
 
 If significant CP violation occurs among the leptons, it could be communicated to the baryons 
 during the early evolution of the universe, thus accounting for the observed baryon number
 asymmetry \cite{lepto}.  At least on a qualitative level, a laboratory discovery of leptonic CP violation
 would put a cosmological observation in accord
 with what we know about low-energy physics.  Making a detailed connection between 
 low-energy observations -- either oscillations or the effects of CP phases in $\beta \beta$-decay
 mass determinations -- and the baryon asymmetry is more difficult.  That connection may
 have to await the invention of a new standard model that explains recent $\nu$ discoveries.
 
 There is a second future concordance test that may be more quantitative, the absolute scale
 of neutrino mass.  Because oscillations test only mass differences, laboratory constraints on
 the neutrino mass scale must come from direct tests in processes like tritium $\beta$-decay
 or indirect tests such as $\beta \beta$-decay (where masses enter in combination with phases).
 The current bounds on neutrino mass
  \begin{equation}
   0.05~\mathrm{eV} \lsim \sum_{i=1}^3 m_\nu(i) \lsim 6.6~\mathrm{eV}
   \end{equation}
  come from the tritium $\beta$-decay limit of 2.2 eV, the possibility of a quasi-degenerate
  neutrino mass spectrum, and the lower bound on mass that can be derived from the
  atmospheric mass difference $\delta m_{23}^2$.  A new tritium experiment now under construction,
  KATRIN \cite{KATRIN}, should push the 2.2 eV bound down to about 0.2 eV, 
  while future $\beta \beta$-decay
  experiments may probe masses at the level of 10-50 meV \cite{50meV}. 
   
   Cosmological analyses -- specifically, constraints from cosmic microwave
   background measurements, large-scale structure surveys, and distance measurements based
   on Type Ia SN observations -- already have lead to the claim \cite{cosmonumass}
   \begin{equation}
   \sum_{i=1}^3 m_\nu(i) \lsim 0.5~\mathrm{eV}.
   \end{equation}
   With anticipated Planck measurements and results from surveys like LSST, this
   upper bound could drop to 50 meV.  
   Thus there is the possibility that cosmological analyses
   will discover neutrino mass through its contributions to dark matter.
   If so, an important ``concordance'' challenge for laboratory expermentalists will be to
   verify that mass by direct measurements.
   
   \section{Laboratory Astrophysics}
   Important progress has been made in laboratory nuclear astrophysics.
   New high-statistics measurements of the reaction $^7$Be(p,$\gamma$) important to the
   pp-chain and solar neutrinos have greatly reduced one of the key uncertainties 
   in SSM $^8$B neutrino flux predictions \cite{junghans, baby},
   \begin{equation}
   S_{17}(0) = 21.4 \pm 0.5 (\mathrm{exp}) \pm 0.06 (\mathrm{theory})~\mathrm{eV-b} .
   \end{equation}
   
   The cross section for $^{14}$N(p,$\gamma$), which controls the CNO rate in main
   sequence stars, was recently remeasured at TUNL \cite{TUNL} and 
   LUNA \cite{LUNA}, Gran Sasso's special-purpose 
   accelerator for nuclear astrophysics.  The LUNA results are
   remarkable and show the potential of underground experiments to
   reduced backgrounds.   The measurements extend to
   center-of-mass energies of 70 keV, the region of the Gamow peak.  Thus not only are the
   measurements precise, but theoretical uncertainties connected with
   extrapolating higher energy measurements to the Gamow peak are eliminated.  The
   resulting S-factor,
   \begin{equation}
   S_{14+1}(0) = 1.61 \pm 0.08~\mathrm{kev-b}
   \end{equation}
   is almost a factor of two below the former best value.  This is a major change, altering for
   example the evolutionary track that stars follow along the red giant branch, delaying 
   He ignition that marks the transition to the asymptotic giant branch.
   Consequences include an increase in the 
   estimated ages of the oldest stars in globular clusters of about 0.8 Gyr, a reduction in the
   flux of CNO solar neutrinos by half, and a delay in the onset of CNO-cycle burning in the
   early massive stars thought to be responsible for reionization.
   
   The success LUNA is enjoying has increased interest in a nuclear astrophysics accelerator
   for DUSEL, the NSF's proposed Deep Underground Science and Engineering Laboratory.
   Designs currently under study include both a conventional light-ion machine and a
   heavy-ion facility for measuring cross sections
   in inverse kinematics \cite{JINA}.  Such high-intensity machines, coupled with advanced detectors, 
   could improve our knowledge of many critical cross sections.
   
   \section{New Data on Neutron Stars}
   Neutron stars (NS) are compact objects, with masses in the range of 1.2-2.0 M$_\odot$,
   radii of
   7-15 km, and magnetic fields that might range to $10^{15}$ Gauss.  NSs can be seen
   through their x-ray and optical emissions.  The rate at which they cool has been determined
   from observations.  These objects are the 
   focus of a great deal of theoretical attention 
   because their structure depends on properties of strongly interacting matter at high density
   and unusual isospin -- issues include the EOS of such matter, the possibility of phase
   changes or coexisting mixed phases, etc.
   
   In the past four years there have been several important observations that promise to
   provide more information on NSs: \\
   $\bullet$ The 1.396 ms (716 Hz) pulsar Terzan5, the
   fastest yet recorded, was identified \cite{Terzan5}.  \\
   $\bullet$  A double-pulsar system PSRJ0737-3039 with known masses was identified \cite{double}. \\
   $\bullet$  A controversial measurement of a large redshift, 0.35, was claimed for H- and He-like 
   Fe lines in x-ray bursts from NS EXO 0748-676 \cite{redshift}.
   
   These results, if confirmed, are important to the goal of understanding the relationship between
   a NS's mass and radius, an important constraint on the nuclear EOS.  It has long been recognized
   that it is difficult to derive radius constraints from isolated NSs due to uncertainties in distance
   determinations and atmospheric modeling.  Stefan-Boltzmann-like
   expressions that relate the NS's radius to the observed flux, surface temperature, and distance
   \cite{LP}
   \begin{equation}
   F_\infty = \sigma T_{eff}^4 \left( {R_\infty \over d} \right)^2
   \end{equation}
   yield results of questionable accuracy.  In contrast, a measurement of the red shift of iron lines
   coming from the surface of a neutron star would, if confirmed, effectively measure
   $G M/R$.  As the system in question is an eclipsing binary where a mass determination should
   be possible, this would determine the radius.
   
   Similarly, milli-second pulsars can place important bounds on NS radii because the
   NS's equatorial velocity -- which would be about 0.25 c for Terzan5 -- cannot 
   exceed the general relativistic analog of Kepler's velocity.  The
   relationship derived by Lattimer and Prakash \cite{LP},
   \begin{equation}
   v_{\mathrm{Kepler}} \sim 1045 \left( {M \over M_\odot} \right)^{1/2} \left( {10~\mathrm{km} \over R} \right)^{3/2}~\mathrm{Hz}
   \end{equation}
   effectively places an upper bound on the radius as a function of $M$, thereby constraining possible
   EOSs.
   
   Alternatively, the double-pulsar system PSRJ0737-3039 with known masses should allow
   observers to determine general relativistic spin-orbit-coupling
   perturbations governed by the star's moment of inertia.  The expected
   accuracy with which the moment of inertia can be determined, given five years of
   patient observations, is about 10\%.
   
   It thus appears that we are about to learn a great deal about NS properties and the 
   nuclear EOS at high density.
   
   There is an important connection between NS properties and the neutrinos produced by
   the SN parent of the NS.  One of the best constraints on $M/R$ might come from integrating
   the neutrino light curve out to very long times.  As 99\% of the energy of a SN is
   emitted in neutrinos -- the origin of this energy is the binding of the neutron star -- such 
   a measurement would determine the general relativistic analog of $GM/R$.  This requires
   good knowledge of the distance to the supernova as well as
   terrestrial detectors capable of measuring the $\nu_e$, $\bar{\nu}_e$, and
   heavy-flavor neutrino fluxes.  As an opportunity to observe a galactic SN may
   arise only once every 30 years, it is important that ``SN watch'' efforts include
   detectors with sensitivities to all flavors and the requisite mass to follow the neutrino
   light curves out to long times, $\gsim$ 20 seconds.
   
   There is also a lovely connection with the parity program at Jefferson Laboratory.  Competing
   nuclear EOSs differ in their predictions for neutron-matter pressures $P_n$ and,
   consequently, NS radii ($\sim P^{1/4}$) because
   of uncertainties in the treatment of the symmetry energy.  The pressure $P_n(\rho_n)$ 
   exerted by neutron-dominated matter is roughly proportional to $dE_{sym}/d \rho_n$, 
   where $\rho_n$ is
   the neutron density \cite{horowitz,steiner}.  
   The skins of heavy, neutron-rich nuclei are another source of neutron-rich matter.
   Parity-nonconserving electron scattering can be used to probe these skins, as the weak 
   charge is effectively the neutron number.  
   Typel and Brown \cite{typel} have argued that isovector nuclear radius depends on $P_n$ through
   the relationship
   \begin{equation}
   \langle r_n^2 \rangle^{1/2} - \langle r_p^2 \rangle^{1/2} \sim P_n(\sim 0.1/f^3).
   \end{equation}
   The density at which $P_n$ is probed, of course, is at or below half nuclear density.  Thus
   there remains the problem of extrapolating to densities important to NSs.
   Nevertheless, an important constraint on the nuclear EOS can be
   obtained in this way.
   
   \section{Conclusions and Remembrances}
   I believe at no time has the field of nuclear astrophysics been more exciting than today.
   The discoveries recently made in neutrino astrophysics have given us our first glimpses of 
   physics beyond the standard model.  The neutrino parameters that have emerged from
   these discoveries are now inspiring a new generation of accelerator neutrino
   experiments to resolve questions about the neutrino mass hierarchy, 
   $\theta_{13}$, and leptonic CP violation.  The results will be important to some
   of the most interesting open questions in astrophysics:
   the origin of matter, the contribution of neutrinos to hot dark
   matter, the mechanism for core-collapse SN.
   
   Laboratory astrophysics will remain important to this field.  LUNA's
   measurement of the controlling reaction of the CNO cycle underscores
   how uncertain our knowledge is of key nuclear reactions governing stellar evolution.
   The field's most important successes -- identification and resolution of
   the solar neutrino problem, the BBN prediction of $\eta$ --
   came from a combination of observation,
   theory, and laboratory measurements.  Today we need laboratory data not only
   for cross sections, but also to constrain the new phenomena associated with
   neutrino masses and mixing.  To understand nucleosynthesis in stars, 
   laboratory measurements have to include both stable and unstable targets.
   
   Large-scale computing has become the third leg of nuclear astrophysics, complementing
   theory and experiments/observations.   It now appears likely that the Type II SN mechanism
   is intrinsically multi-dimensional, and that successful models of the collapse will require
   the coupling of magneto-hydrodynamics, shock-wave propagation, neutrino transport
   of energy and lepton number, and rotation.   Success in building a ``standard
   model'' of SN could produce benefits comparable from those derived from the SSM.   
   The supernova environment tests
   unique aspects of neutrino physics ($\theta_{13}$, the MSW potential of neutrinos), 
   nucleosynthesis (the explosive conditions necessary for the r-process, the neutrino
   fluences required for the $\nu$-process), and gravity (gravitational waves tagged
   by the SN's optical signal).
   
   Observations of millsecond pulsars like Terzan5 and the double-pulsar system PSRJ0737-3039
   suggest that stringent constraints on the properties of
   dense, neutron-rich nuclear matter
   will soon be available.   NSs could provide crucial constraints
   on models of the nuclear EOS at high density.
   
   This year is also one for remembering three colleagues who did so much to build this field,
   Hans Bethe, John Bahcall, and Vijay Pandharipande.   Almost every topic in my talk today
   has roots in the work of these pioneers.   Hans is truly the father of nuclear astrophysics,
   as it was his efforts to understand hydrogen burning as the mechanism for main-sequence
   stellar evolution that built the bridge from the stars to the laboratory.  Because of John's
   tireless efforts, the solar neutrino problem was finally recognized as a triumph for 
   quantitative nuclear astrophysics and as the first low-energy manifestation of the
   physics that lies beyond the standard model.   Vijay pioneered techniques to
   derive the properties of nuclei and nuclear matter directly from NN and NNN forces,
   including properties of nature's largest nucleus, the neutron star.   We will miss them all.
   
   This work was supported in part by the Office of Nuclear Physics, U.S. Department of 
   Energy, and by the DOE's SCIDAC program.
   


\end{document}